\documentclass[useAMS,usenatbib]{mn2e}  

\usepackage{dsfont}
\usepackage{amsmath}
\usepackage{amssymb}
\usepackage{graphicx}
\usepackage{verbatim}
\usepackage{natbib}
\usepackage{eucal}
\usepackage{calligra}
\usepackage{url}

\usepackage{amsfonts}
\usepackage[usenames,dvipsnames]{color}
\usepackage[normalem]{ulem}
\usepackage[T1]{fontenc}

\DeclareMathAlphabet{\mathscr}{OT1}{pzc}%
                                 {m}{it}
     
\newcommand{\simgt}{\lower.5ex\hbox{$\; \buildrel > \over \sim \;$}}
\newcommand{\simlt}{\lower.5ex\hbox{$\; \buildrel < \over \sim \;$}}
                            
\newcommand{\mnras}{MNRAS}

\newcommand{\apj}{ApJ}

\newcommand{\aap}{A\&A}

\newcommand{\nat}{Nature (London)}

\newcommand{\be}{\begin{equation}}
\newcommand{\ee}{\end{equation}}
\newcommand{\bes}{\begin{equation*}}
\newcommand{\ees}{\end{equation*}}
\newcommand{\bea}{\begin{eqnarray}}
\newcommand{\eea}{\end{eqnarray}}
\newcommand{\beas}{\begin{eqnarray*}}
\newcommand{\eeas}{\end{eqnarray*}}

\newcommand{\rpair}{R_{\rm pair}}
\newcommand{\mpch}{\;{\rm Mpc}/h}

\newcommand{\zl}{z_{\rm L}}
\newcommand{\zs}{z_{\rm s}}

\def\ave#1{\left\langle #1 \right\rangle}

\begin{document}

\title[Stacked Filament Lensing]
  {Detection of Stacked Filament Lensing Between SDSS Luminous Red Galaxies}
\author[J. Clampitt et al.]
  {Joseph~Clampitt$^{1}$\thanks{E-Mail: clampitt@sas.upenn.edu},
  Hironao Miyatake$^{2,3,4}$, Bhuvnesh~Jain$^{1}$, Masahiro~Takada$^3$\\  
  $^1$Department of Physics and Astronomy, University of Pennsylvania,
  209 S. 33rd St., Philadelphia, PA 19104, USA\\
$^2$Department of Astrophysical Sciences, Princeton University,
Peyton Hall, Princeton NJ 08544, USA\\
$^3$Kavli Institute for the Physics and Mathematics of the Universe
(Kavli IPMU, WPI),
The University of Tokyo, Chiba 277-8583, Japan\\
$^4$Jet Propulsion Laboratory, California Institute of Technology, Pasadena, CA 91109
}

\maketitle

\begin{abstract}
We search for the lensing signal of massive filaments between 135,000 pairs of Luminous Red Galaxies (LRGs) from the Sloan Digital Sky Survey.
We develop a new estimator that cleanly removes the much larger shear signal of the neighboring LRG halos, relying only on the assumption of spherical symmetry.
We consider two models: a ``thick'' filament model constructed from ray-tracing simulations for $\Lambda$CDM model, and a ``thin'' filament model which models the filament by a string of halos along the line connecting the two LRGs.
We show that the filament lensing signal is in nice agreement with the thick simulation filament, while strongly disfavoring  the thin model.
The magnitude of the lensing shear due to the filament is below $10^{-4}$.
Employing the likelihood ratio test, we find a 4.5$\sigma$ significance for the detection of the filament lensing signal, corresponding to a null hypothesis fluctuation probability of $3 \times 10^{-6}$.
We also carried out several null tests to verify that the residual shear signal from neighboring LRGs and other shear systematics are minimized.
\end{abstract}


\section{Introduction}

One of the most striking features of $N$-body simulations for
$\Lambda$CDM structure formation scenario
is the network
of filaments into which dark matter particles arrange themselves. Some
attempts to quantify this network have been made \citep{s2011, cwj2014}. Other
work has attempted to study the largest filaments, those between close
pairs of large dark matter halos \citep{ckc2005}. Such filaments are
likely the easiest to identify in data, e.g., \citet{zdm2013} look for
overdensities in the galaxy distribution between close pairs of galaxy
clusters. However, since filaments include both dark and luminous
matter, weak lensing techniques are  useful to understand the entire
structure: \citet{dwc2012} and \citet{jjk2012} both identify single filaments by focusing on a
weak lensing analysis of individual cluster pairs.

In this study we measure the weak lensing signal of filaments
between stacked Luminous Red Galaxy (LRG) pairs in Sloan Digital Sky Survey (SDSS) data.
The mass distribution and therefore weak lensing shear in the neighborhood of LRG pairs is dominated by the massive halos themselves.  Methods which aim at filament detection, e.g., \citet{mm2013}, may have a degeneracy with the signal from these nearby halos. In the face of this degeneracy, we construct an estimator of the lensing signal which removes the shear due to these halos, assuming only that they are spherically symmetric. We will show that this technique is sufficient to obtain a detection, and some physical implications on filament size and shape can be extracted by comparison to filament models. Systematic errors which are expected to be spherically symmetric with respect to the halos, such as intrinsic alignments, are nulled simultaneously.

Other work has attempted to estimate the feasibility of weak lensing stacked filament detection. \citet{mm2013} make optimistic choices for survey parameters and find that $\sim 2 - 4\sigma$ detections are possible for single clusters but state that their method has difficulties in application to stacked filament detection. 
In another study \citet{mkm2010} use lens and source redshifts that make
their lensing strength a factor of 2 greater than ours, and a galaxy
number density at least a factor of 30 higher. The lower mass limit of
their stacked clusters is $M_{200} = 4 \times 10^{14} M_\odot / h$, much
larger than the dark matter halos associated with our LRGs. With these
parameters, they estimate that $\sim 20$ cluster pairs are necessary to
obtain a detection.

Filaments can also be characterized using the language of higher-order
 correlations. In this case, one would describe the filament as the part
 of the matter-matter-matter three point function in the neighborhood of the
 halos forming a cluster pair. A detection of the halo-halo-matter
 3-point function around such cluster pairs was made
 using the Red Cluster Survey \citep{sps2008}. More recently \citet{simonetal:13} used CFHTLens survey to measure the galaxy-galaxy-shear correlation function and
attempted to measure the average mass distribution around galaxies. 
 This measurement was done by subtracting off the two point contribution of the lensing signal. As these authors discovered, the three-point signal peaks at the cluster locations. However, for our purposes of identifying filaments, such a location of the three point function's peak makes the technique of two-point subtraction unsatisfactory. Just as our nulling estimator removes two-point contributions which are spherically symmetric about the halo centers, it also removes any three-point contribution which is centered on these points.  
 
We use a different approach to these authors.
We use data from the SDSS which is shallower, but covers 8,000 square degrees. This allows us to use $\sim$ 100,000 pairs of LRGs and obtain a significant detection with a new estimator of filament lensing.
 
The outline of this paper is as follows: section \ref{sect:method} describes the basic nulling technique for removing spherically symmetric components. In section \ref{sect:data} the LRG pair catalog and background source shear catalog used in this work are described. Section \ref{sect:theory} describes our mock LRG catalog and ray-tracing simulations, as well as an alternative thin-filament model. In section \ref{sect:results} we present our main results, including the filament measurement from data and simulations, null tests, and the likelihood ratio statistic. Finally, section \ref{sect:discussion} discusses the implications of our results as well as directions for future work.

Throughout this work we use cosmological parameters $\Omega_{\rm m} = 0.3$, $\Omega_{\Lambda} = 0.7$, and $\sigma_8 = 0.83$.

\section{Measurement Technique}
\label{sect:method}

In this section, we describe the nulling technique for spherically
symmetric components, which includes most of the two-point signal and
the peak of the three-point signal.

\subsection{Nulling spherical components}

We bin the data in such a way as to null the shear signal from any
spherically symmetric source at the location of either member of the
halo pair. To first order, such halos are expected to follow a spherically-symmetric NFW
density distribution \citep{nfw1997} when stacked. However our technique is not dependent on the precise shape of the halo profile, only on its spherical symmetry.
We note that halo anisotropy which might be
preferentially aligned with the inter-pair direction would not be
nulled by the following procedure, but its small contribution is treated in Appendix~\ref{app:ellip}.

\begin{figure}
\centering
\resizebox{80mm}{!}{\includegraphics{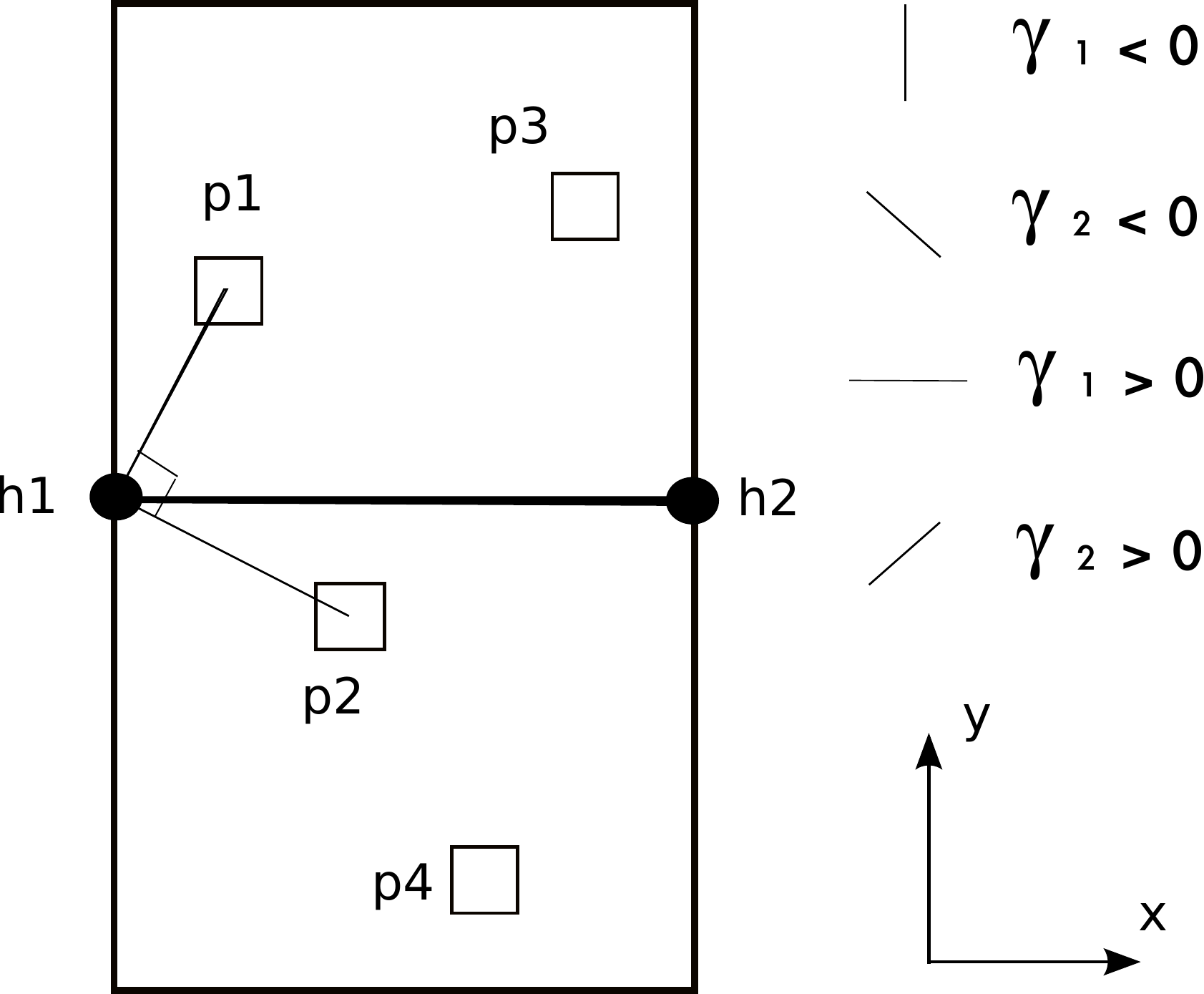}}
\caption{Combining data in points p1-p4, the average shear signal from
 spherical halos h1 and h2 is zero. 
The point ``p2'' is the counter point of ``p1'' with respect to
 halo ``h1'', while the points ``p3'' and ``p4'' are the counterparts of
 p2 and p1 with respect to halo ``h2'', respectively.
This nulling method only works when
 all shears are measured relative to the fixed Cartesian
coordinate system on the sky (as indicated at bottom right).
Our convention for the sign of the two shear components is given by the $\gamma_1$ and $\gamma_2$ whiskers. }
\label{fig:points}
\end{figure}

\begin{figure}
\centering
\resizebox{45mm}{!}{\includegraphics{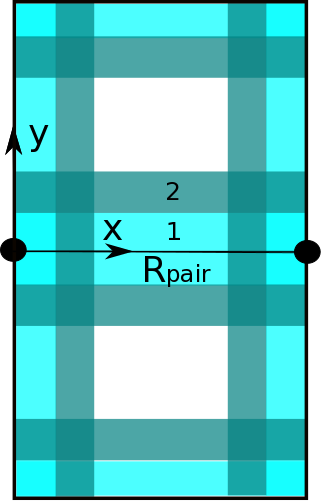}}
\caption{The lensing measurement (cross-component null test) is performed by combining
 all background shears' $\gamma_1$ ($\gamma_2$) components in bins, such
 as the pictured bins 1 and 2. The hypothesized filament should lie along the line connecting the two LRGs
 denoted by bold points.
}
\label{fig:nulling}
\end{figure}

First consider just one spherically symmetric halo, h1, as pictured in
Fig.~\ref{fig:points}. Pick any point p1 nearby. Draw another point p2
which is (i) 90 degrees away from p1 with respect to the halo, and (ii)
at the same distance from the halo as p1. The tangential shears
$\gamma_t$ from these points add, 
while the cross component
$\gamma_\times$ is zero. This is the standard galaxy-galaxy lensing
measurement. But if the shear components at p1 and p2 are measured with
respect to a fixed
coordinate system on the sky, they average out to zero. We denote 
the shear components relative to
this fixed Cartesian
coordinate system $\gamma_1$ and $\gamma_2$. As shown in
Fig.~\ref{fig:points}, we choose this coordinate system such that
the $x$-axis is along the line connecting h1 and h2 (the two
halos) for each halo pair,
$\gamma_1 < 0$ is perpendicular to the $x$-axis, and $\gamma_1 > 0$ is parallel.

Now add a second halo, h2. We need to null the h2 shear signal in both p1 and p2 as well. To do so, rotate both points by 90 degrees about h2 to make points p3 and p4. By construction, the average $\gamma_1$ and $\gamma_2$ shear signal measured at these four points has no contribution from a spherical halo at h2. Furthermore, one can check that rotating p3 by 90 degrees about h1 brings it into p4, so that this set of four points is null with respect to both halos.

Such sets of four points are the building blocks for a number of
possible binning schemes which attempt to null the spherically-symmetric
halo signal.
Note that any set of bins which exploit this property will necessarily mix scales relative to the hypothesized filament. However, since the most likely location for an inter-halo filament is on the line connecting the halo pair, we choose bins which will minimize this mixing of scales. The background shears are separated into bands that run parallel or perpendicular to the filament direction in Fig.~\ref{fig:nulling}. The first two such bins are numbered on the figure. This binning scheme also exploits the expected symmetries about the center of the filament, in both horizontal and vertical directions. To verify that a bin does indeed fulfill the conditions for nulling the spherical signal mentioned above, imagine rotating the part of the bin above the $R_{\rm pair}$ line about either halo, and see that it goes into the same colored bin in the region below the line. Note also that each background source is counted twice due to the overlap between different bins. This means a naive shape noise accounting of errors would underestimate the noise by a factor $\sqrt{2}$.

In what follows, we describe our measurement procedures of filament lensing.
For the halos h1 and h2 in Fig.~\ref{fig:points}, we use pairs of LRGs (see Sec.~\ref{sect:pair} for details), most of which are central galaxies of dark matter halos.
Implicitly, the above discussion assumes that the local coordinate system is defined such that both LRGs lie along the $x$-axis.
Thus, we begin by rotating the \texttt{RA} and \texttt{DEC} coordinates of each LRG pair and all nearby background galaxies (those within the boundaries pictured in Fig.~\ref{fig:nulling}) so that the LRG pair always lies along the $x$-axis.
It follows that the $y$-coordinate of background sources then denotes the distance from the line between the LRG pair.
In addition, the source catalog has shear components $(\gamma'_1, \gamma'_2)$ which are also defined with respect to the \texttt{RA}, \texttt{DEC} coordinate system: we simultaneously rotate these into ($\gamma_1, \gamma_2$) components defined with respect to the above $(x,y)$ system.

Then, following the method in \citet{msb2013}, our lensing observable is given by the shear of background sources $\gamma_k$ at the pixel ($x,y$) times the lensing weight $\Sigma_{\rm crit}$.
The precise estimator is
\begin{equation}
\Delta\Sigma_{k}(x,y;\zl)=\frac{\sum_j 
\left[
w_j \left(\ave{\Sigma_{\rm
  crit}^{-1}}_j(\zl)\right)^{-1} \gamma_{k}(\vec{x}_j)
\right]
}{\sum_j w_j} \, ,
\label{eq:dSigma}
\end{equation}
where 
the summation $\sum_j$ runs over all the background galaxies
in the pixel ($x,y$), around all the LRG pairs,
the indices $k = 1,2$ denote the two components of shear, 
and the weight for the $j$-th galaxy is given by
\be \label{eq:weight}
w_j = \frac{\left[
\ave{\Sigma_{{\rm crit}}^{-1}}_j(\zl)\right]^{2}}
{\sigma_{\rm shape}^2 + \sigma_{{\rm meas},j}^2}.
\ee
We use $\sigma_{\rm shape} = 0.32$ for
the typical intrinsic ellipticities  and 
$\sigma_{{\rm meas},j}$ denotes
measurement noise on each background
galaxy. 
Again notice that, when computing the average shear field, we use the same
coordinate system for each LRG pair: taking one LRG at the coordinate
origin and taking the $x$-axis to along the line connecting two LRGs as
pictured in Fig.~\ref{fig:points}.
$\ave{\Sigma_{\rm crit}^{-1}}_j$ is the lensing critical density for the
$j$-th source galaxy, computed by taking 
into account the photometric redshift uncertainty:
\be
\ave{\Sigma_{{\rm crit}}^{-1}}_j(\zl)
= \int_0^\infty\! {\rm d}\zs \Sigma_{\rm crit}^{-1} (\zl, \zs) P_j(\zs),
\ee
where $\zl$ is the redshift of the LRG pair and $P_j(\zs)$
is the probability distribution of photometric redshift
 for the $j$-th galaxy.
Note that $\Sigma_{\rm crit}^{-1}(\zl,\zs)$ is computed as a function of lens
and source redshifts for the assumed cosmology as
\be
\Sigma_{\rm crit}^{-1} (\zl, \zs) = \frac{4\pi G}{c^2} \frac{D_A(\zl) D_A(\zl,\zs)}{D_A(\zs)}
\ee
and we set $\Sigma_{\rm
crit}^{-1}(\zl,\zs)=0$ for $\zs<\zl$ in the computation. 

To increase statistics, we will measure the stacked 
weak lensing signal of
filaments as a function of distance $y$ from the line connecting the two
LRGs. Based on our nulling method in
Fig.~\ref{fig:points}, each ``p1'' point at distance $y$ has its
counterparts with coordinate values
\begin{equation}
 {\rm p1}(x,y) \rightarrow 
\left\{
{\rm p2}(y, -x), {\rm p3}(1-x,1-y), {\rm p4}(1-y,x-1)\right\},
\end{equation}
where we set the first LRG position ``h1'' as the coordinate center
$(x,y)=(0,0)$, and we have used the units of $R_{\rm pair}=1$ for
convenience. 
Hence we employ the following estimator of filament lensing
signal for the $a$-th distance bin, $y_a$, in Fig.~\ref{fig:nulling}:
\begin{eqnarray}
 \widehat{\Delta\Sigma_k^{\rm fil}}(y_a)
&\equiv&
\!\!\!
\sum_{0<x_b<0.5}\!\!
\left[
\Delta\Sigma_k(x_b,y_a)
+
\Delta\Sigma_k(y_a,-x_b)\right.\nonumber\\
&&\hspace{-5em}
+
\Delta\Sigma_k(1-x_b,1-y_a)
+
\Delta\Sigma_k(1-y_a,x_b-1)\nonumber\\
&&\hspace{-5em}
+
\Delta\Sigma_k(x_b,-y_a)
+
\Delta\Sigma_k(y_a, x_b)\nonumber\\
&&\hspace{-5em}
\left.
+
\Delta\Sigma_k(1-x_b, y_a-1)
+
\Delta\Sigma_k(1-y_a, 1-x_b)
\right],
\label{eq:est_signal}
\end{eqnarray}
where $\Delta\Sigma_k(x,y)$ denotes the $k$-th component of average shear
at the position ($x,y$) (see Eq.~\ref{eq:dSigma}, but note that the sum in the denominator of Eq.~\ref{eq:dSigma} runs over all lens-source pairs in the bin when plugged into Eq.~\ref{eq:est_signal}).
We use the notation $\Delta\Sigma^{\rm fil}$ to denote the shear field caused by the gravitational tidal field due to a filament, with dimensions of the surface mass density, but it should not be confused with the surface mass density used in galaxy-galaxy lensing.
The summation is over the $x$-coordinate of the sources, and the summation
range is
confined to $0<x_b<0.5$ in order to avoid a double counting of
the same background galaxies 
in the different quads of points p1, \dots, p4.
Note however that the above binning does put each galaxy in two
different bins. The third and fourth lines of Eq.~(\ref{eq:est_signal})
exploit the symmetry about the line joining the LRG pair, by letting
$\Delta\Sigma_k(x,y) \rightarrow \Delta\Sigma_k(x,-y)$. Putting each
galaxy in two bins in this way does add to our covariance between bins
(as we will later discuss around Fig.~\ref{fig:covariance}), but even so there is a gain in information. This is because when a galaxy is put in, say, bin 1 it is averaged together with a different set of galaxies compared to when it is placed in bin 2.

In the preceding discussion on nulling the LRG lensing signal, we implicitly assumed that sufficiently many background sources would be found in each pixel $(x,y)$ so that none of p1-4 (Fig.~\ref{fig:points}) are empty of sources.
For example, if there were sources located at p1, p2, and p3, but not p4, the LRG signal would not be nulled.
Here we check the assumption that the stacked background source density is suffiently large so as to guarantee that each group of four pixels has a nulled LRG signal.
First, note that we are performing a stacked lensing measurement with $\sim 135,000$ LRG pairs (see Sec.~\ref{sect:pair}), so the total number of sources falling in a given pixel can be estimated as the number for the typical LRG pair times 135,000.
Since we will use 8 bins (see Sec.~\ref{sect:results}) covering $0 < y < R_{\rm pair}/2$ and $R_{\rm pair} \ge 6 \mpch$, our smallest pixel is $3/8 \mpch$.
At the typical lens redshift of $z \sim 0.3$ this physical scale corresponds to an angular scale of $\sim 2$ arcminutes.
The SDSS background source density (see Sec.~\ref{sect:sources}) is $\sim 0.25 \, {\rm arcmin}^{-2}$ at this redshift, so for a single LRG pair, we would find $\sim 1$ galaxy in a single pixel.
Then, for the stacked lensing measurement we have approximately $100,000$ sources in each pixel, easily satisfying the assumption that the stacked, spherically-symmetric LRG shear will be nulled on average.

\subsection{Halo ellipticity}

The nulling technique has the extra benefit of mostly removing contributions
from halo ellipticity, expected to point along the line joining the LRG
pair. The ellipticity-direction cross-correlation of \citet{lsp2008} has
shown that simulated dark matter halos tend to point towards other halos
in their vicinity. While the intrinsic alignment of LRGs has been
measured at a less significant level, the smallness of the intrinsic alignment of the galaxy ellipticity 
is more likely due to misalignment of the light and mass profiles
\citep{ojl2009,clampittjain:15}, rather than the lack of alignment between neighboring massive halos. But if we let the virial radii of these halos be $\Delta \leq 1 \mpch$ and the pair separation be $\rpair \geq 6 \mpch$, then the ratio of these $\Delta / \rpair$ is a small quantity, and we show in Appendix~\ref{app:ellip} that contributions to the signal are highly suppressed as this ratio gets smaller.

\subsection{Jackknife Realizations}

We perform the measurement and all null tests by first dividing up the survey area of 8,000 sq. deg. into 134 approximately equal area regions. We then measure each quantity multiple times with each region omitted in turn to make $N=134$ jackknife realizations. The covariance of the measurement \citep{nbg2009} is given by
\begin{flalign} \label{eq:cov}
C [\Delta\Sigma_i^{\rm fil}, & \Delta\Sigma_j^{\rm fil}] = \frac{(N-1)}{N}
& 
\nonumber \\
& 
\times \sum\limits_{k=1}^N \left[(\Delta\Sigma_i^{\rm fil})^{k} - \overline{\Delta\Sigma^{\rm fil}_i}\right]
\left[(\Delta\Sigma_j^{\rm fil})^{k} - \overline{\Delta\Sigma^{\rm fil}_j}\right]
\end{flalign}
where the mean value is
\be \label{eq:avg}
\overline{\Delta\Sigma^{\rm fil}_i} = 
\frac{1}{N}
\sum\limits_{k=1}^N (\Delta\Sigma^{\rm fil}_i)^k\, ,
\ee
and 
$(\Delta\Sigma^{\rm fil}_i)^k$
denotes the measurement from the $k$-th realization and the $i$-th
spatial bin. The covariance is measured for both components of shear;
for clarity we do not denote the separate shear components in Eqs.~\ref{eq:cov} and \ref{eq:avg}.

\section{Data}
\label{sect:data}

\subsection{Pair catalog}
\label{sect:pair}

\begin{figure}
\centering
\resizebox{83mm}{!}{\includegraphics{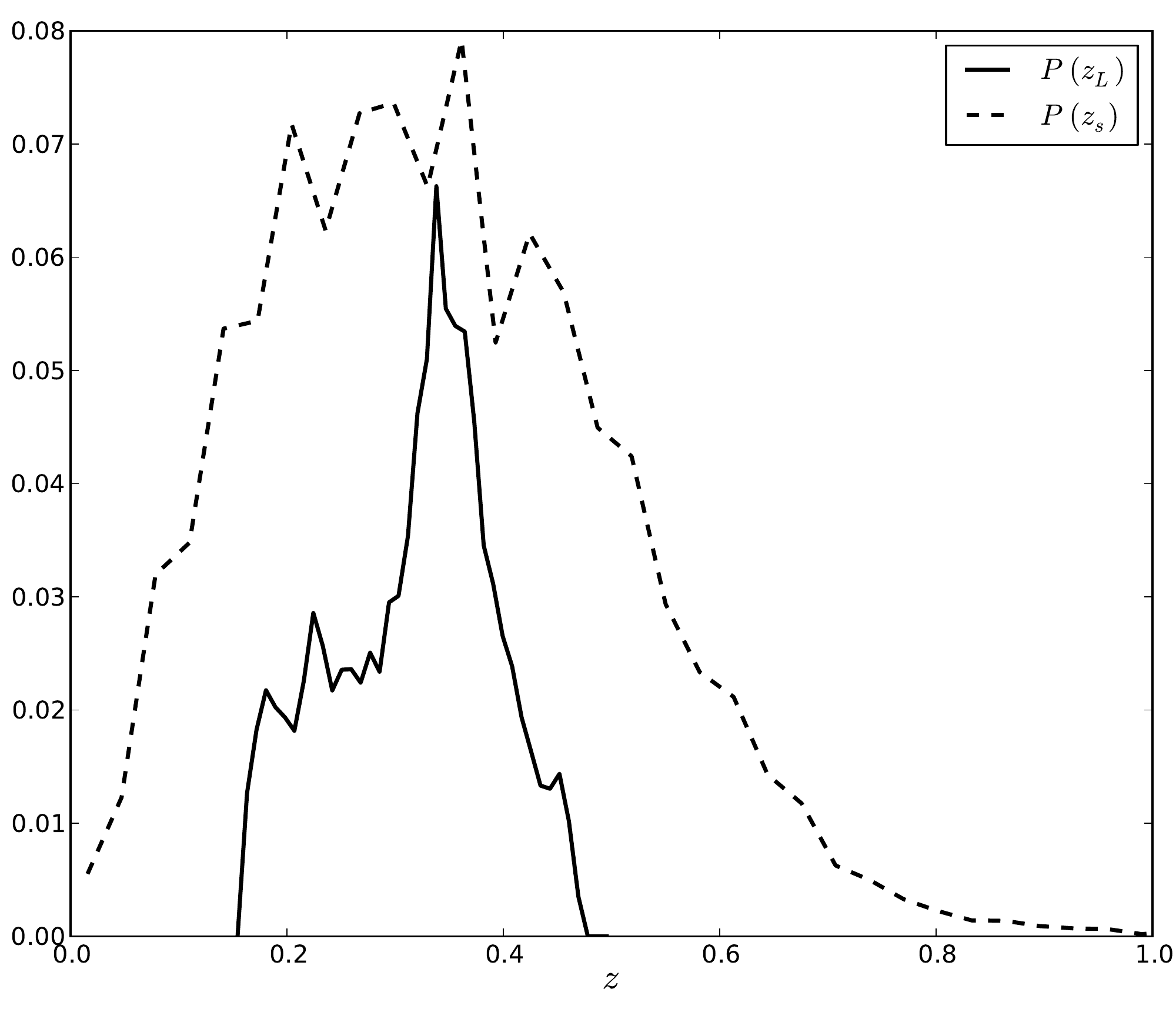}}
\caption{The redshift distribution of LRG pairs used as lenses (solid line) and background sources (dashed line).
}
\label{fig:pair}
\end{figure}

We use the SDSS DR7-Full LRG catalog of \citet{kbs2010}, which contains
105,831 LRGs between $0.16 < z < 0.47$. The sky coverage is
approximately 8,000 sq. deg. The pair catalog is constructed by choosing
each LRG in turn, and finding all neighboring LRGs within a cylinder of
physical (or proper) radius $14 \mpch$ and physical line-of-sight distance $\pm 6
\mpch$. The redshift distribution of our pairs is in the left panel of
Fig.~\ref{fig:pair}. The distribution in line-of-sight distance differences between the pair members is
roughly uniform.
The cutoff of $|\Delta r_{\rm los}| < 6 \mpch$ corresponds roughly to a
redshift separation of $\Delta z < 0.004$ between pairs.
Note that this line-of-sight separation assumes the
LRG velocity is only due to Hubble flow; in other words, 
the redshift difference can arise from the difference of
line-of-sight peculiar velocities ($\Delta v=1200$ km/s for $\Delta r_{\rm los} = 6 \mpch$) 
even if the two LRGs are in the
same distance. This is the so-called redshift space distortion (RSD),
and we will discuss the effect of RSD on our weak lensing measurements.

We obtain $\sim 135,000$ pairs with the separation cutoffs given above: since
each LRG can be a member of multiple pairs, this is slightly more than the number of objects
as in the original LRG catalog. With $\rpair$ defined to be the physical projected
separation between the LRGs, for pairs between $6 \mpch < \rpair < 14
\mpch$ we have a distribution $P(\rpair)$ which grows very slightly with
$\rpair$. The virial radii of these halos are $\sim 0.5 - 1.0 \mpch$, so our selection of objects with $\rpair \geq 6 \mpch$ ensures that these LRGs live in different dark matter halos. We have checked that the measurement is insensitive to the choice of physical vs. comoving distances.

\begin{figure}
\centering
\resizebox{83mm}{!}{\includegraphics{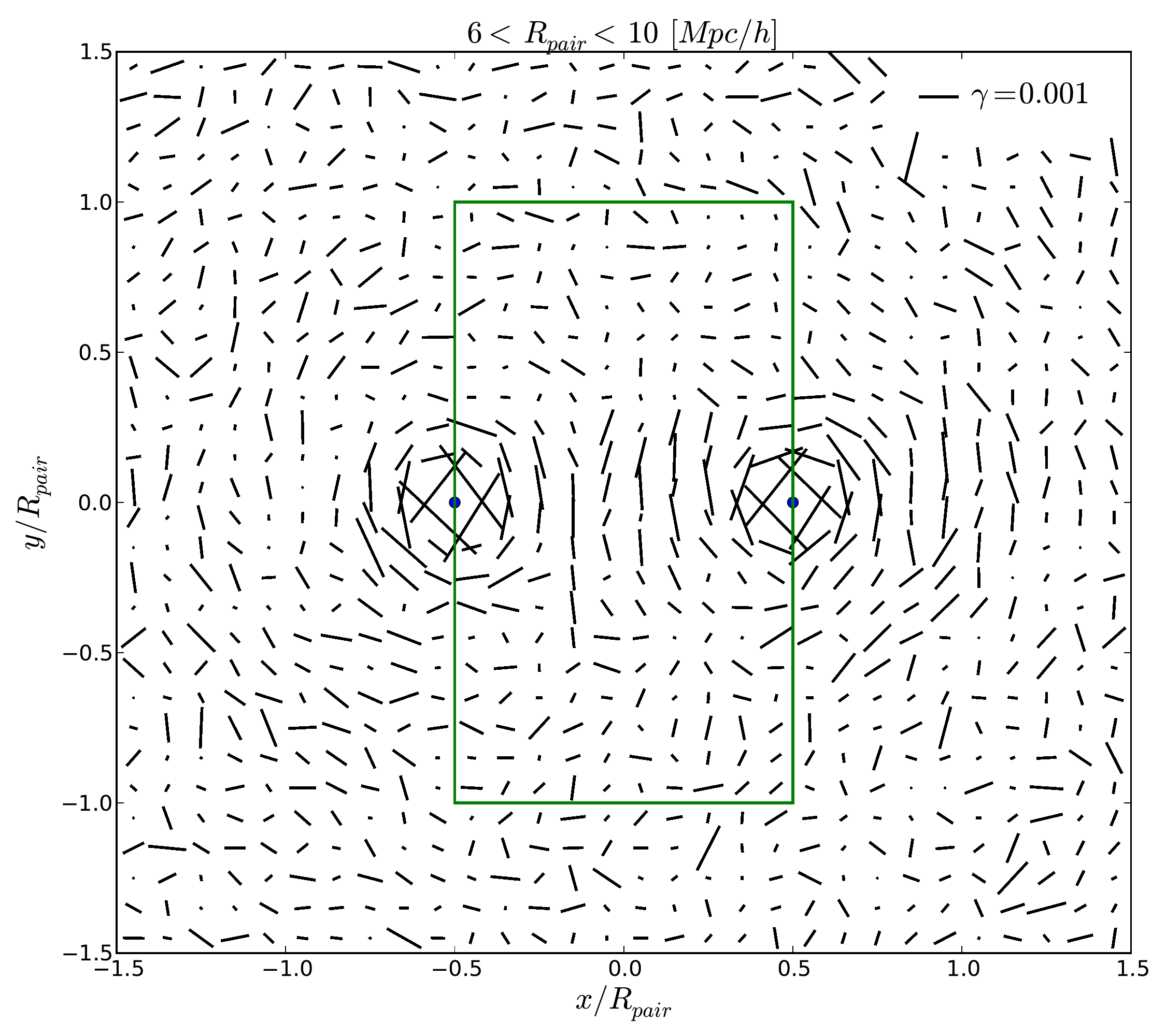}}
\caption{The stacked shear field for our smallest separation bin,
 $6 \mpch < \, \rpair < \, 10 \mpch$, obtained by stacking the
 background galaxy ellipticities in the same Cartesian coordinate system
 around each LRG pair region (see Fig.~\ref{fig:points} and Eq.~\ref{eq:dSigma}).
The tangential shear signal of the LRG halos is clearly visible.
The green box shows the filament measurement region of Fig.~\ref{fig:nulling}.
 }
\label{fig:data-whisker}
\end{figure}

In Fig.~\ref{fig:data-whisker} we show the stacked
shear whiskers for the smallest $\rpair$ bin; each lens-source pair is optimally weighted as in Eqs.~(\ref{eq:dSigma}) and (\ref{eq:weight}), and we convert back to $\gamma$ by assuming fiducial redshifts $\zl = 0.25$ and $\zs = 0.4$. The tangential shear signal around each member of the LRG pair is clearly visible. The nearest whisker to each LRG has magnitude $\approx 0.003$. Note that due to the large distance between whiskers (0.1$\rpair \sim 1 \mpch$) even the closest ones to each halo are far from the center at $\sim R_{\rm vir} /2$. The dominance of the LRG halos in these fields motivates our use of the nulling scheme to isolate the relatively tiny filament lensing signal.

\begin{figure*}
\centering \resizebox{180mm}{!}{\includegraphics{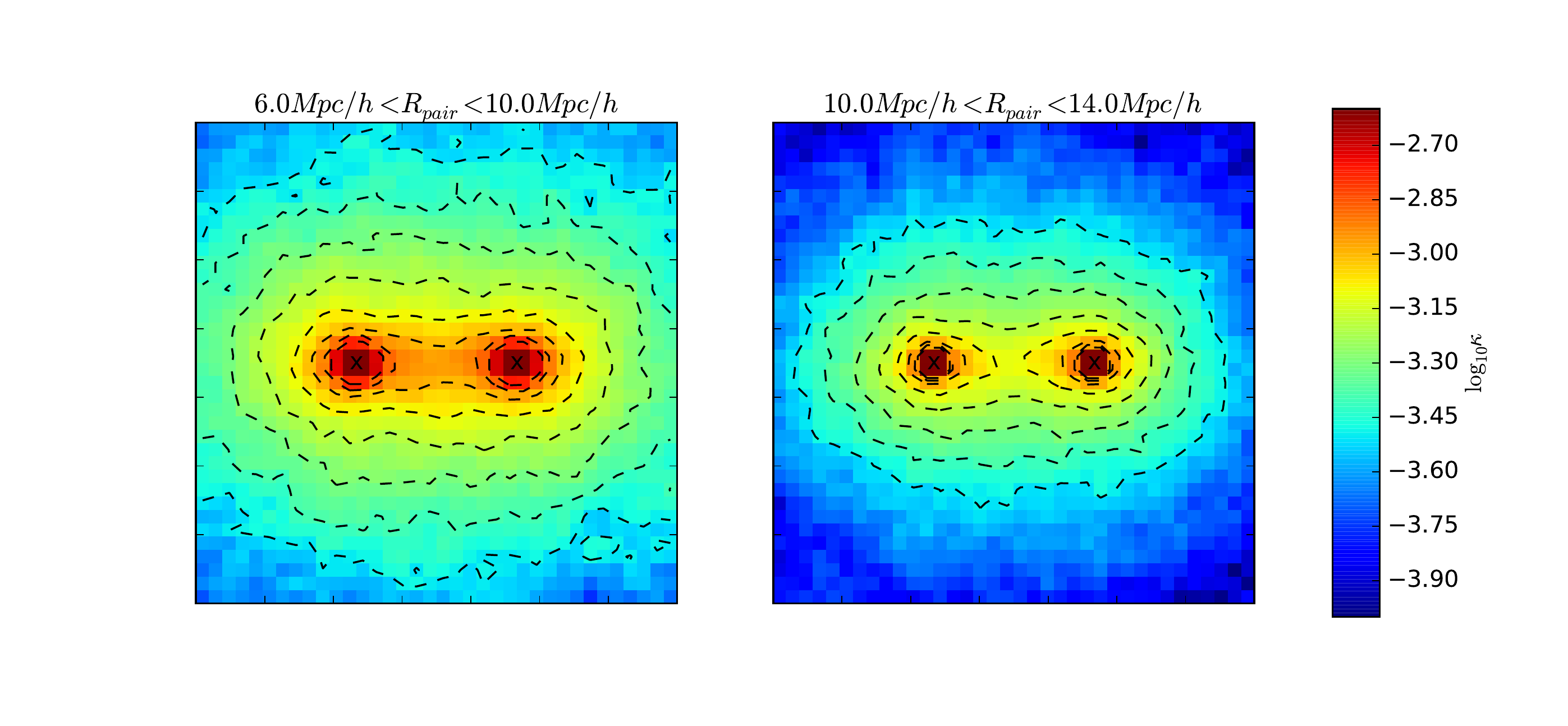}}
\caption{The $\kappa$ map of stacked LRG-like halo pairs in ray-tracing
simulations. The left (right) panel shows pairs with the projected
separation between $6-10~\mpch$ ($10-14~\mpch$). The positions of halos are marked by ``$\times$''. The
contours show $\kappa$ in logarithmic scale with the interval of $\Delta
\log_{10} \kappa = 0.1$. The mass excess, i.e., filament, is clearly
seen between the pairs.  Both panels show that the filament is thicker (a few~$\mpch$) than
the size of halos virial radii ($\sim 0.5-1 \mpch$).}  \label{fig:sim_kappa}
\end{figure*}

\subsection{Background source catalog}
\label{sect:sources}

The shear catalog is composed of 34.5 million sources, and is nearly
identical to that used in \citet{sjs2009}. The source redshift
distribution is shown in Fig.~\ref{fig:pair},
and is obtained by
stacking the posterior probability distribution of photometric
redshift for each source, $P(\zs)$.
While the peak of this source catalog is approximately at the same redshift as the peak of our LRG pairs, $z \sim 0.35$, the source distribution has a substantial tail extending out to higher redshifts. For further details of the shear catalog, see \citet{sjs2009}.

\section{Theory: Thick- and thin-filament models}
\label{sect:theory}

We compare the measurement to the following two models, which generally
predict ``thick'' or ``thin'' filaments, respectively: (i) a model
obtained based on ray-tracing simulations or
(ii) a one-dimensional string of less massive NFW halos (a collection of NFW halos making up the 1D filament).

\begin{figure*}
\centering
\resizebox{160mm}{!}{\includegraphics{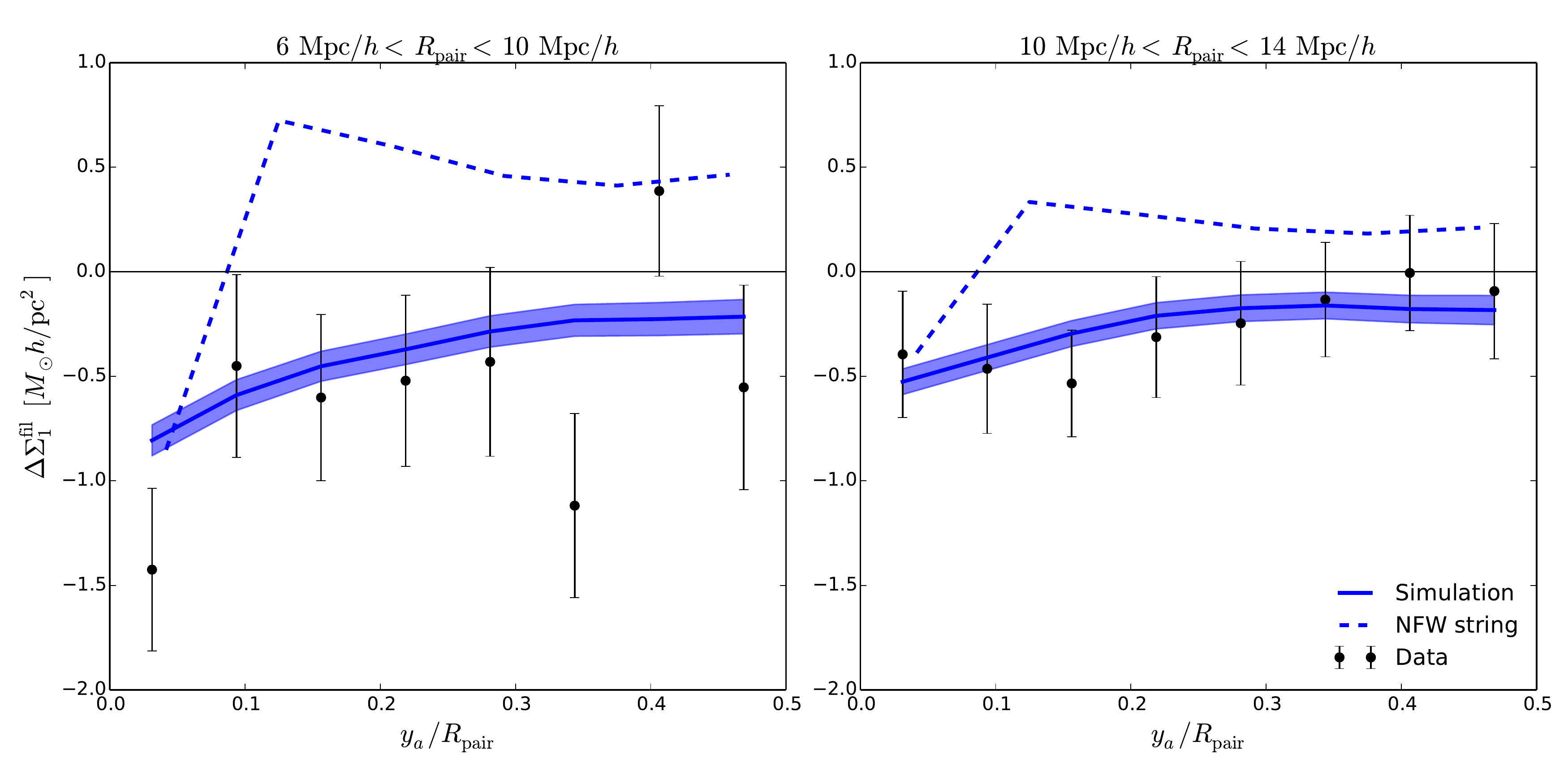}}
\caption{Filament measurement $\Delta\Sigma^{\rm fil}_1$ (black points)
for the closer set of pairs {\it (left panel)} and more widely
separated pairs {\it (right panel)}. We compare the measurement to two
theoretical models, the simulation model (bold blue line) and NFW string
(dashed blue).  The band around the blue line shows variations
in the simulation predictions for three mock catalogs each of which has
a SDSS-like area ($8000$ sq. degrees). Hence this indicates the sample
 variance in the model prediction (see text for details).  The shape of the simulation
prediction, a thick filament, is supported by the data, while the
thin-filament NFW string is ruled out along with the null hypothesis.}
\label{fig:measure}
\end{figure*}

\subsection{Thick-filament model from ray-tracing simulations}
Here we use a ray-tracing simulation to make a prediction for the weak
lensing signal between LRG pairs. We first select LRG-like halos in the
simulations, and then make pairs and carry out the same measurement as
in Section \ref{sect:method}.

\subsubsection{Mock catalog of LRG pairs based on ray-tracing simulations}
\label{sect:mocks}

To model the weak lensing signal of filaments between LRG pairs,
we use ray-tracing simulations in
\cite{satoetal:09}\footnote{\url{http://www.a.phys.nagoya-u.ac.jp/~masanori/HSC/}
or available from M. Takada based upon request.}. Note that they
used slightly different cosmological parameters in the simulations;
$\Omega_m=0.238$, $\Omega_\Lambda=0.762$, and $\sigma_8=0.76$.
\footnote{The $\sigma_8$ value is smaller than what the recent CMB
experiments have implied, $\sigma_8\simeq 0.8$. Since the filament
lensing signal would scale with $\sigma_8$ as $\propto (\sigma_8)^4$ in the weakly
nonlinear regime, based on the picture of the three-point correlation
function, we might underestimate the lensing signal by $\simeq 20\%$ if
the universe follows $\sigma_8=0.8$.}
In brief, the simulations were generated based on
the algorithm in \cite{HamanaMellier:01}, using $N$-body simulation
outputs of large-scale structure for a $\Lambda$CDM cosmology.  In this
paper, we use the 1000 realizations of simulated lensing fields for
source redshift $z_s=0.6$, where each realization has an area of
$5\times 5=25$ square degrees. Hence the ray-tracing simulation
effectively covers an area of 25,000 sq. degrees in total.  The lensing
fields, convergence and shear, are provided in the format of $2048^2$
pixels for each realization.
The catalog of halos is also available for each ray-tracing realization.
The halos were identified from the $N$-body simulation output used in
each lens plane,
using the friends-of-friends (FoF) algorithm with linking length of 0.2
in units of interparticle spacing. The catalog contains the FoF mass,
angular position and redshift for each halo, where the halo position was
taken from the center-of-mass of FoF member particles, and the redshift
was computed from the distance in the light-cone simulation.

To build a mock catalog of LRGs, we employ the halo occupation
distribution (HOD) method to populate hypothetical LRGs into the
simulated halos in the range of $0.16\le z\le 0.47$, as in the
measurements. In this paper we use the HOD model in
\cite{reidspergel:2009}. Note that \citet{reidspergel:2009} employed the
spherical overdensity (SO) halo finder with $\Delta=200\bar{\rho}_m$,
and the halo masses of SO and FoF methods might differ by about 10\%
\citep{tinkeretal:2008}.
The HOD consists of the two contributions, the mean occupation
distribution for central and satellite galaxies given as a function of
host halo mass: $\langle N \rangle(M)=\langle N_{\rm cen}\rangle(M)
[1+\langle N_{\rm sat}\rangle(M)]$. 
For a central LRG, we randomly select halos according to the HOD
probability at the halo mass, which effectively selects all halos at the
high mass end where $\langle N_{\rm cen}\rangle \simeq 1$.
For satellite LRGs, assuming the Poisson distribution with
the mean given by $\langle N_{\rm cen}(M)\rangle\langle N_{\rm
sat}\rangle(M)$, we generate a random number from the Poisson
distribution, take the integer number $\hat{N}_{\rm sat}$, and then
populate $\hat{N}_{\rm sat}$ LRG(s) into each
halo. Note that we do not populate
any satellite LRGs into a halo if
the halo does not host a central LRG.
In this step, we ignored the distribution of satellite LRG(s)
inside each host halo: i.e. put the LRG(s) at the halo center.  The
length scale of filaments in which we are interested in is much longer
than the size of LRG host halo, so our treatment should be a good
approximation.
Our mock catalog properly accounts for the fact
that halos which host multiple LRGs inside are counted multiple times
in the lensing measurement.

As discussed in \S~\ref{sect:pair}, the RSD due to the peculiar
velocities of LRGs might affect a selection of LRG pairs. 
However, the halo catalog does not contain the velocity
information. Hence we add the RSD displacement to each LRG's position in
the mock catalog by assuming a Gaussian distribution with
$\sigma_v=500~$km/s (corresponding to a typical radial displacement of
$\simeq 5.7~{\rm Mpc}/h$ at $z=0.35$ in the comoving units).

Once the mock catalog of LRGs in each simulation realization is
constructed, we implement the same measurement method (the selection of
LRG pairs and the measurement of filament lensing).
Fig.~\ref{fig:sim_kappa} shows the stacked maps of lensing convergence
field ($\kappa$) between the LRG pairs, obtained from the 1000
realizations. The ray-tracing simulations indicate a ``thick'' filament
between the LRG pairs, which has a width of a few \mpch.

\subsection{Thin-filament model}
\label{sect:thin}

Here we consider a ``thin'' string-of-halos model as an independent
model from the simulations.
For this simple model, we use a 1D line of NFW \citep{nfw1997} halos as in \citet{mm2013}.
The model has just two parameters: $M_{\rm fil}$, the total mass in the
string of halos, and $N_{\rm fil}$, the number of halos in the
string. Each halo is given a mass $M_{\rm halo} = M_{\rm fil} / N_{\rm
fil}$, and different halos are equally-spaced along the string between two LRGs.
However, we have checked that the prediction is not very sensitive to the choice of $N_{\rm fil}$.
To generate predictions for this model,
we calculate the shear profile at any given point by adding up the contribution for each
halo in the string, with each halo's contribution calculated according to the exact solution for NFW shear given by \citet{wb2000}.
The overall shear amplitude depends on the total mass $M_{\rm fil}$.
This model generally predicts the shear pattern that
is parallel to the string (i.e. $\gamma_1>0$), at the distance $y\simgt 1 \mpch$.
Since the small NFW halos making up the filament sit exactly on the line between the LRG pair, we call this the ``thin-filament'' model.
In the following section, we use $M_{\rm fil} = 2\times 10^{14}
h^{-1}M_\odot$ and $N_{\rm fil} = 20$, so that the mass per halo is
$10^{13} M_{\odot}/h$. For all the halos we assumed $c=7$ for the
concentration parameter, although the dependence is very weak.
The mass was chosen to give a magnitude roughly equal to the simulation prediction.
However, since this model does not give the correct sign for most measurement bins (see the following section), the actual amplitude is not important.
\begin{figure}
\label{fig:error_jackknife}
\centering
 \resizebox{80mm}{!}{\includegraphics{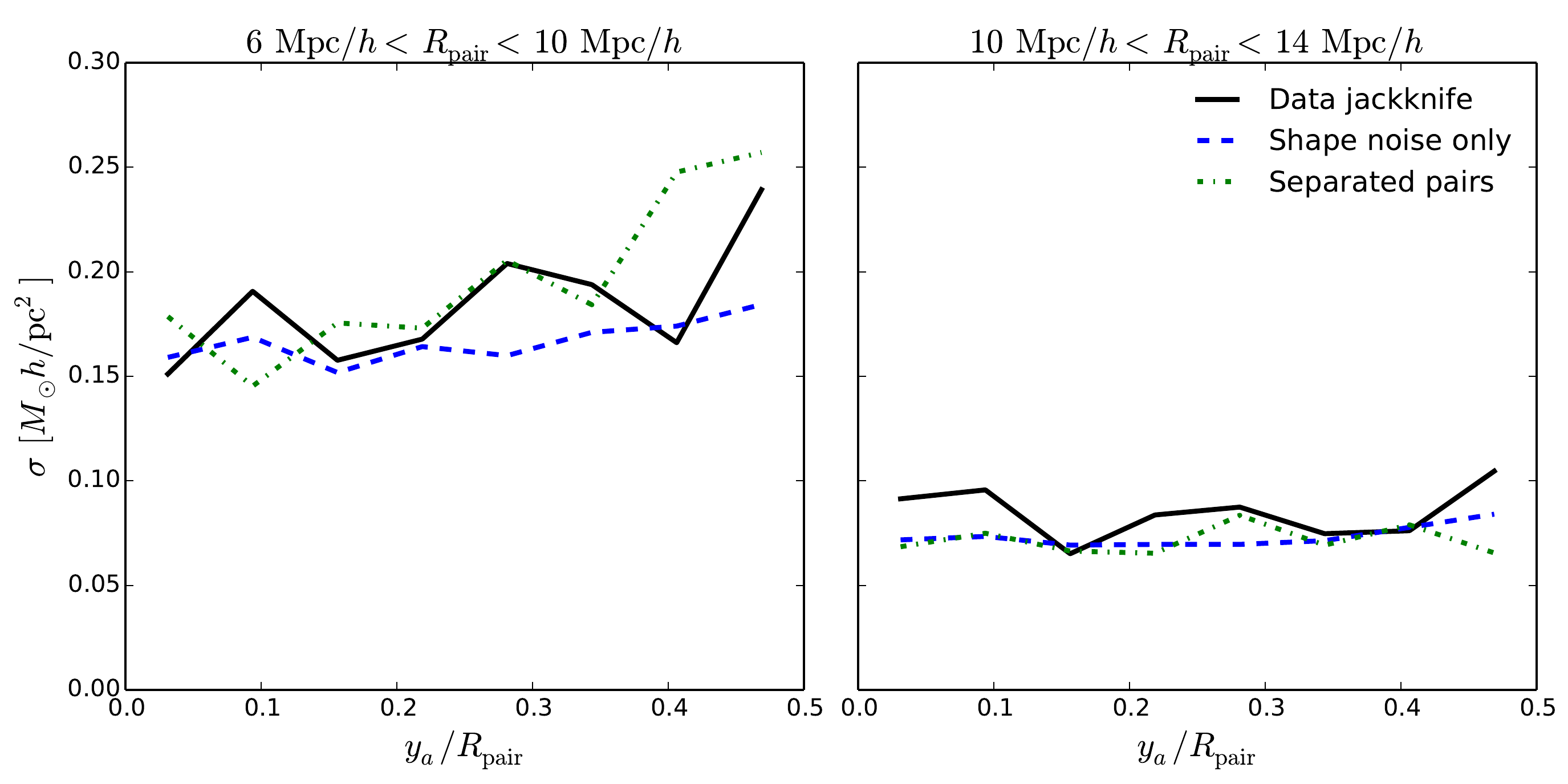}}
 \caption{Different estimates of the error bars of filament
 lensing measurements, which are the square root of the diagonal
 components of the covariance matrix obtained from the jackknife method
 on the data.  ``Data jackknife'' denotes the estimate from the LRG
 pairs, ``Shape noise only'' is the estimate from the LRG pair regions,
 but after randomly rotating orientation of each background galaxy
 ellipticity, and ``Separated pairs'' denotes the estimate from the LRG
 pairs, which have the same projected separation as our main LRG pair
 sample, but have line-of-sight separation of $100 \mpch < \Delta r_{\rm
 los} < 120 \mpch$ (see text for details). These three results are very
 similar, indicating that shape noise dominates the error bars.}
\end{figure}

\begin{figure*}
\centering
\resizebox{180mm}{!}{\includegraphics{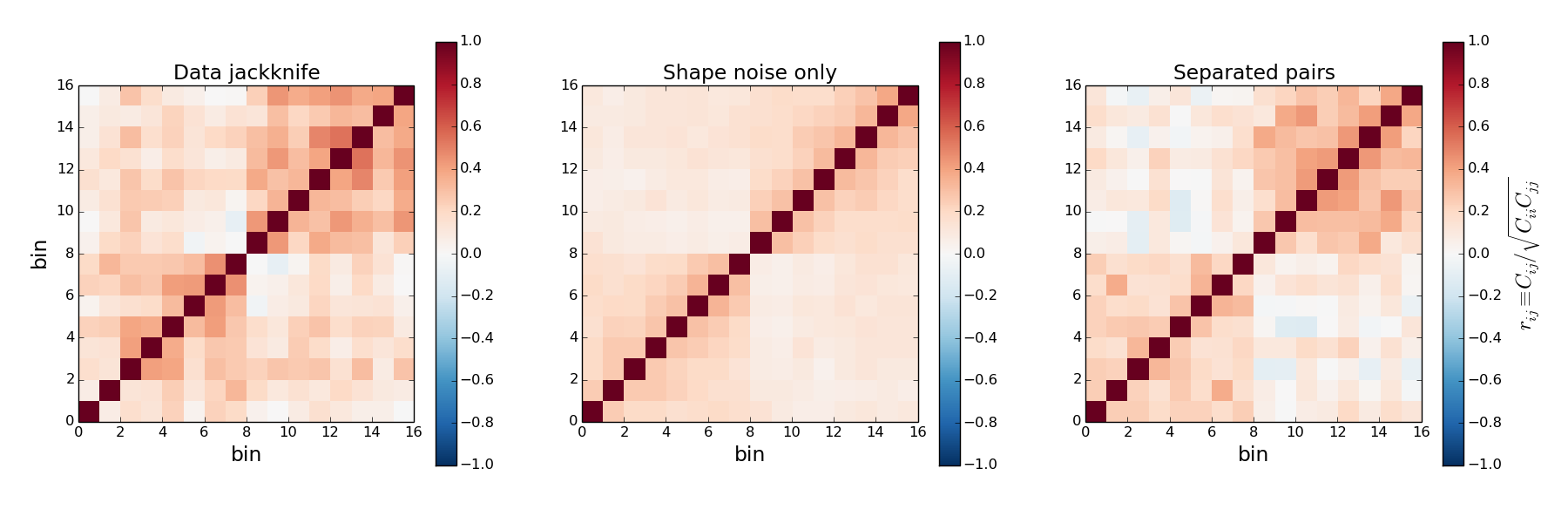}}
\caption{({\it left panel}): The normalized covariance matrix of $\Delta\Sigma^{\rm fil}_1$ as measured from the data using three variants of the jackknife method.
Bins 1-8 and 9-16 represent the two samples of LRGs with $6-10\mpch$ and $10-14\mpch$ separation, respectively.
({\it middle panel}): The same, but showing covariance from shape noise alone, obtained by applying a random rotation to source galaxies before performing the measurement.
({\it right panel}): The same, but for the Separated pair test from the data.
}
\label{fig:covariance}
\end{figure*}

\section{Results}
\label{sect:results}

\subsection{Measurement}

In Fig.~\ref{fig:measure} we show the filament estimator Eq.~(\ref{eq:est_signal}) applied to the data.
The magnitude of the signal for the sample with pair separation $\rpair
= 6-10 \mpch$ is $\sim 0.7 \, h M_\odot/{\rm pc}^2$, with little variation between bins.
This is roughly two orders of magnitude smaller than the shear signal from the LRGs themselves, depending on scale \citep{msb2013}.
The shear signal is diminished between LRG pairs with larger separations of $10-14\mpch$, as expected if such pairs are less likely on average to be connected by filaments \citep{ckc2005}.
Recalling our sign convention in Fig.~\ref{fig:points}, the negative sign of $\Delta\Sigma_1^{\rm fil}$ corresponds to shear aligned perpendicular to the filament axis.
This is the expected direction of the shear at small radii within filaments, as predicted by \citet{mm2013}.
That we measure negative signal in all bins is not inconsistent with the fact that far from the filament the shear direction should be tangential and thus change sign.
Our binning mixes scales (Fig.~\ref{fig:nulling}) in order to null the spherical halo signal, and each bin includes background sources which are both near to and far from the line between the LRG pairs.
Due to this mixing of scales, it is not easy to guess the filament shape by eye based on the data points alone.

Nonetheless, by comparing to specific models we can still make quantitative statements about the filament properties.
In Fig.~\ref{fig:measure} we compare the simulation and thin-filament models to the data.
With a sign-flip relative to the data, the thin-filament model is easily ruled out, but the
thicker filament predicted by the simulations matches the data very well.
In the following section, we quantify this agreement between the simulations and data, and in addition describe our method for calculating the departure of the data from the Null hypothesis of no filament.
Looking at the shape of the thin-filament prediction, it is clear that if we can rule out the Null hypothesis then this thin-filament model is ruled out as well.

\subsection{Likelihood ratio test}
\label{sec:lrtest}

To evaluate the significance of the filament lensing signals in
Fig.~\ref{fig:measure}, we need to estimate the error covariance
matrix. The covariance matrix arises from two contributions: shape noise
and sample variance. The former arises from an average of
intrinsic ellipticities over a finite number of source galaxies. The
latter arises from variations of filament lensing as well as the
projection effect due to lensing contributions from mass distribution
at different redshifts, but along the same line-of-sight to the LRG
pairs. Since our method mixes different scales due to the nulling of
spherical halo signal, it is not clear which contribution to the
covariance is dominant.

We study the covariance matrix using different jackknife
methods on the data. First, we estimate the covariance matrix based on
the standard method, applying the jackknife method to the real LRG
pairs (Eq.~\ref{eq:cov}). Secondly, we estimate the covariance for the shape noise alone;
we first randomly rotate the orientation of each source galaxy ellipticity,
repeat the filament lensing measurement in Eq.~(\ref{eq:est_signal}),
and then estimate the covariance from the jackknife method. The random
orientation erases the lensing signals from both the filament and the
projection effect. We used 20 realizations to obtain a well-converged
covariance estimate. 
Note that in this method we keep fixed the positions of
source galaxies, and therefore used the same configurations of triplets
of LRG pairs and source galaxies in the covariance estimation. Thirdly,
the covariance is estimated from the jackknife method using the ``separated'' LRG
pairs, which are selected such that the two LRGs have
the same projected separation on the sky as the true LRG
pairs, but have a line-of-sight separation of $100 \mpch < \Delta r_{\rm
los} < 120 \mpch$ (see also \S~\ref{sect:null} for details). The
separated LRG pairs are unlikely to have filaments between the two LRGs, due to
the large three-dimensional separation, but will include similar lensing contributions
from the LRG halos and the projection effect. Comparing these three
covariance matrices reveals the relative contributions to the covariance
from the shape noise, the filaments and the projection effect.

The diagonal components of the covariance matrices are shown in Fig.~\ref{fig:error_jackknife}.
The three results look very similar, indicating that the shape noise is the dominant source of the error bars.
Similarly, Fig.~\ref{fig:covariance} compares the off-diagonal
correlation coefficients of the covariance matrices, $r_{ij} \equiv
C_{ij} / \sqrt{C_{ii} C_{jj}}$, estimated from the three jackknife
methods above. The three results again look similar. It should be noted
that the shape noise alone causes non-vanishing off-diagonal components,
because the filament lensing measurements use the same background
galaxies multiple times. With the results in
Figs.~\ref{fig:error_jackknife} and \ref{fig:covariance}, we conclude
that the shape noise is the dominant source of the error covariance, and
there is no significant contribution of the sample variance.
This conclusion is also justified by the mock catalogs of LRG pairs.
Since we have the ray-tracing simulation data of 25,000 sq. degrees in
total (see \S~\ref{sect:mocks}), we generated 3 SDSS-like mocks each of which
has an area of 8,000 sq. degrees as in the SDSS data. The band around the bold solid line in
Fig.~\ref{fig:measure} denotes the variations in the model predictions
among the 3 mocks, which show the sample variance contribution. The
figure clearly shows that the sample variance is very small compared to
the scatter between different bins.
Hence in the following we use the covariance estimated from the
jackknife method on the data (labelled ``Data jackknife'' in
the figures.)

We now move on to an assessment of the significance of the filament
lensing measurement. To do this, we employ a ``simple-vs-simple''
likelihood ratio test to attempt to rule out the Null hypothesis that
there is no excess mass extending between the LRGs, in favor of the
Simulation hypothesis that the mass distribution looks like that in our
simulated LRG pair catalogs.  The likelihood ratio is the ratio of the
null likelihood, ${\cal L}_0$, to the simulation likelihood,
${\cal L}_{\rm s}$:
\be
\frac{{\cal L}_0}{{\cal L}_{\rm s}} \propto \frac{e^{-\chi_0^2 / 2}}{e^{-\chi_{\rm
s}^2 / 2}} \, ,
\ee
where
\begin{eqnarray}
 &&\chi_{\rm 0}^2 = \sum_{ij} d_i\left(
				  {\bf C}^{-1}
				 \right)_{ij} d_j, \nonumber\\
 &&\chi_{\rm s}^2 = \sum_{ij} (d_i - d^{\rm m}_i)
\left({\bf C}^{-1}\right)_{ij}
  (d_j - d^{\rm m}_j) \, ,
\end{eqnarray}
${\bf C}$ is the covariance matrix estimated from the LRG pairs in
Figs.~\ref{fig:error_jackknife} and \ref{fig:covariance}, ${\bf
C}^{-1}$ is the inverse of the matrix, $d_i$ denotes the central value
of the measurement at the $i$-th bin, and
$d^{\rm m}_i$ is the simulation
prediction at each bin (the solid lines in Fig.~\ref{fig:measure}).
As our test statistic $T$ we use twice the
natural logarithm of the likelihood ratio, dropping an irrelevant
constant from the normalization,
\begin{equation}
T \equiv \chi_{\rm s}^2 - \chi_0^2 \, . \label{eq:T}
\end{equation}

\begin{figure}
\centering
\resizebox{80mm}{!}{\includegraphics{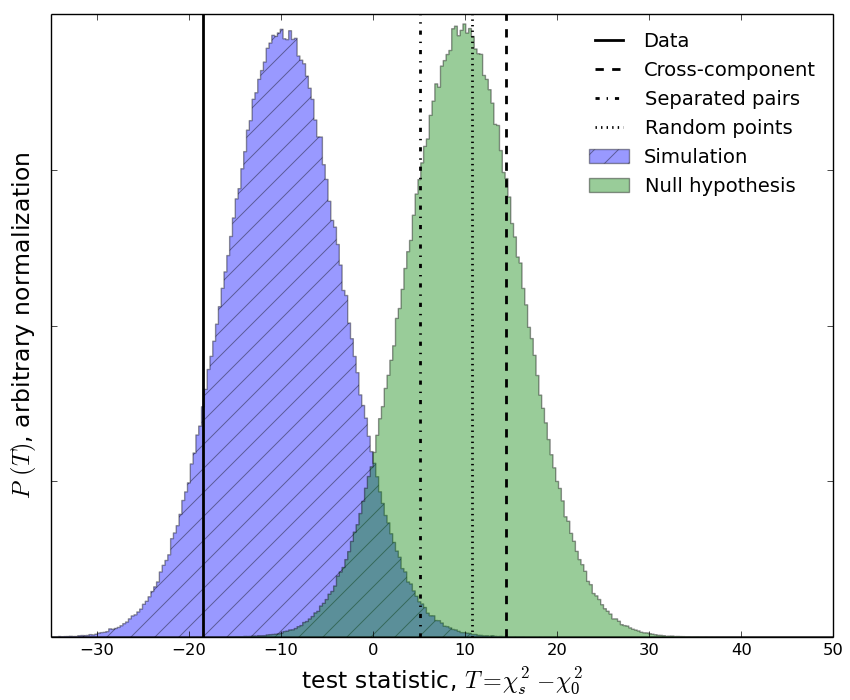}}
\caption{The comparison of the simulation model (blue hatched histogram) with the null hypothesis (green solid histogram) is carried out using a likelihood ratio test.
The filament measurement from data (vertical solid line) can be easily produced if the true mass distribution is similar to that found in the simulation.
In contrast, the null hypothesis of no filament produces a test
 statistic more extreme than the observed measurement in only $\sim 3$
 out of $10^6$ Monte-Carlo samples, corresponding to a $4.5\sigma$
 detection.
 On the other hand, the three null tests, ``Cross-component'',
 ``Separated pairs'' and ``Random points'' (see \S~\ref{sect:null} for
 details) are well within the distribution of null hypothesis.
}
\label{fig:lrtest}
\end{figure}

To make a quantitative comparison of the Null and Simulation
hypotheses, we generate distributions using ``fake'' data vectors as
follows. Assuming a multi-variate Gaussian distribution obeying either
${\cal L}_0\propto \exp(-\chi^2_0/2)$ or ${\cal L}_{\rm s}\propto
\exp(-\chi^2_{\rm s}/2)$, we generate Monte Carlo realizations of the
fake data vector $\hat{d}_i$.
Since we used the same covariance matrix in both $\chi^2_0$ and
$\chi^2_{\rm s}$, the difference between the distributions is due to the
expectation central values: $\langle\hat{d}_i\rangle=0$ and $\langle \hat{d}_i\rangle = d^{\rm m}_i$ for the null and simulation predictions, respectively.
Fig.~\ref{fig:lrtest} shows the
distribution of $T$ values (Eq.~\ref{eq:T}) for the Monte Carlo
realizations of Null and Simulation hypotheses.
For example, a ``typical'' value of the test statistic drawn at the peak of
the Simulation hypothesis histogram falls in the tail of the Null
histogram, and vice-versa.

The results shown in Fig.~\ref{fig:lrtest} show that the actual value of the test statistic $T$ calculated from the data falls near the peak of the Simulation histogram.
In contrast, the data is not typical if the true model is given by the Null hypothesis.
The p-value describing the probability that the Null hypothesis would produce a test statistic more extreme than the data is found to be $0.000003$, allowing us to rule out the Null hypothesis at the $4.5\sigma$ confidence level.
The values of the test statistic for the three null tests (see Sec.~\ref{sect:null}) fall within $1\sigma$ of the peak of the Null histogram, indicating these tests pass.

\subsection{Null tests: Separated pairs, Cross-component, and Random points}
\label{sect:null}

\begin{figure*}
\centering
\resizebox{140mm}{!}{\includegraphics{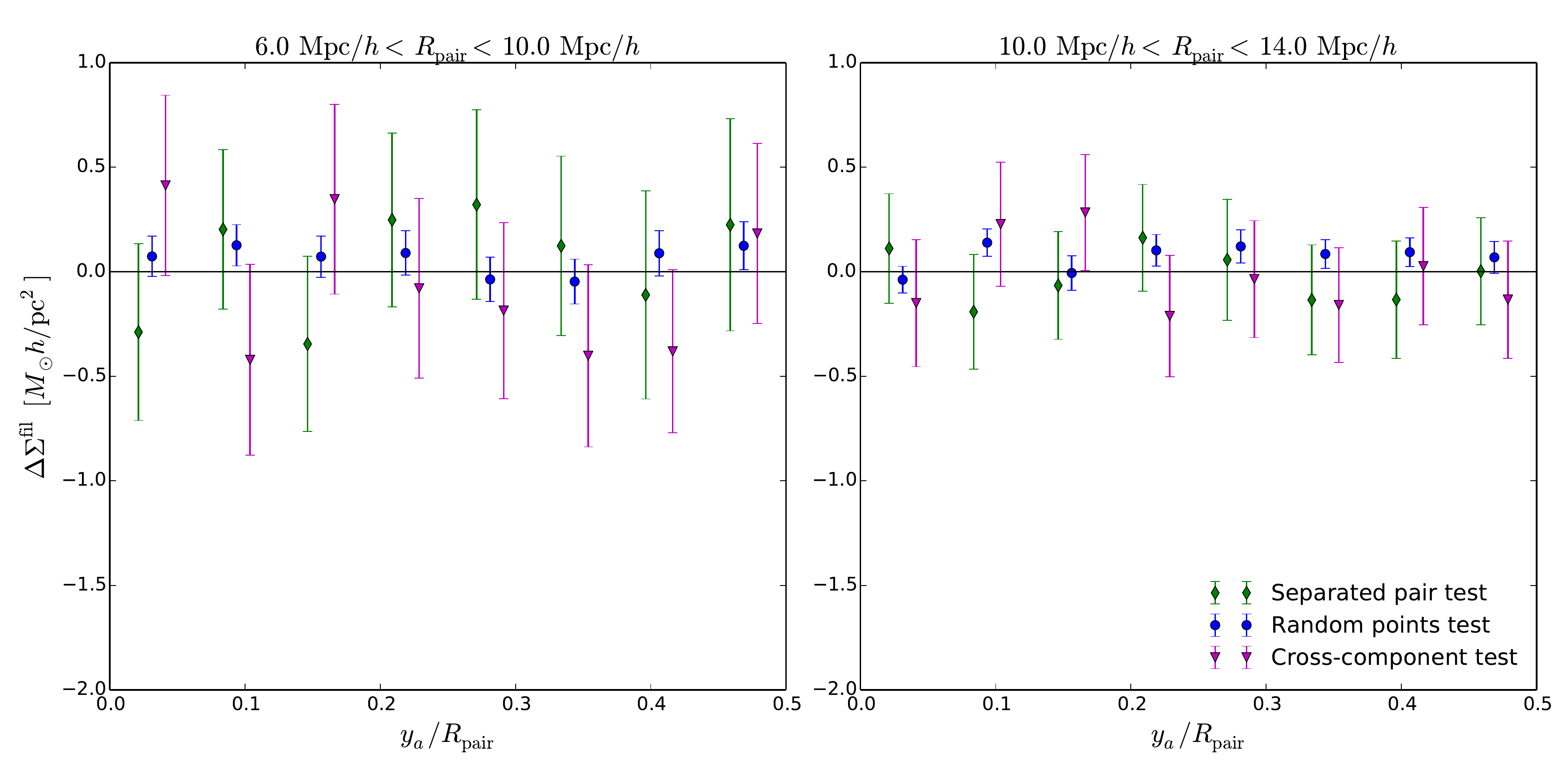}}
\caption{Same as Fig.~\ref{fig:measure}, but showing the results of three null tests (as labelled in the legend). The separated pair test, random points test, and cross-component test are all consistent with the null hypothesis. Note that the null result for the separated pair test shows that our estimator does successfully null the spherically symmetric signal from the LRG halos.}
\label{fig:null-measure}
\end{figure*}

In order to validate the measurement in the previous sections, we perform three null tests on the data.
For all null tests, we repeat the measurement of our Eq.~(\ref{eq:est_signal}) estimator for $\Delta\Sigma^{\rm fil}_1$ using the same jackknife regions.
First, the Separated pair test involves using two LRGs at the ``h1'' and ``h2'' positions of Fig.~\ref{fig:points}, but with line-of-sight separation $100 \mpch < \Delta r_{\rm los} < 120 \mpch$. The 3D distance of such pairs is so large that we expect no excess mass to build up between them. For the lens redshift $z_{\rm L}$, we use the average of the two LRG redshifts. The result is shown in Fig.~\ref{fig:null-measure} (green diamonds) and is consistent with the Null hypothesis (Fig.~\ref{fig:lrtest}). This test shows that the spherically symmetric shear signal from both LRGs in the measurement is truly nulled, as claimed.

This null test measurement has a further use in verifying one of our approximations in Sec.~\ref{sec:lrtest}.
We assumed that the data covariance is a fair approximation to the Null hypothesis covariance, which should include all sources of noise which would appear if the Null hypothesis were true.
This includes shape noise, as well as residual noise from the uncancelled part of the LRG halo signal and other mass distributions which are not completely nulled by our estimator, Eq.~(\ref{eq:est_signal}).
It should not include noise from variations in the purported filament itself, which would not be present given the Null hypothesis.
We find the Separated pair jackknife covariance differs only slightly from the filament covariance itself (Fig.~\ref{fig:covariance}), and the detection significance shifts by less than $1\sigma$ when using this covariance for the Null hypothesis.

Second, as in tangential shear measurements, where the cross-component of shear rotated by $45^\circ$ has no first-order contribution from gravitational lensing, our cross-component (the $\Delta\Sigma^{\rm fil}_2$ component of Eq.~\ref{eq:est_signal}) has no contribution from a filament. This statement holds as long as the stacked mass distribution around the LRG pairs has reflection symmetry about the line joining the pairs. For such a mass distribution, in the Cartesian coordinate system of Fig.~\ref{fig:points}, $\gamma_2(y) = - \gamma_2(-y)$. Since background sources at $y$ are always put in the same bin with sources at $-y$, (see Fig.~\ref{fig:nulling}), $\Delta\Sigma^{\rm fil}_2 = 0$ on average. This is what we find in Fig.~\ref{fig:null-measure}, where the magenta triangles show the result of this null test. Again the result for the test statistic in Fig.~\ref{fig:lrtest} is consistent with the Null hypothesis.

For the Random points test, we repeat the measurement on $\sim 10$ times as many random points with the same distribution in $z$ and $R_{\rm pair}$ as the pair catalog.
The random positions and orientations used in this test should stack individual halos and filaments such that the final mass distribution is isotropic, and thus nulled by our procedure.
The result is shown in Fig.~\ref{fig:null-measure} (blue circles).
Like the others this test passes, as shown in Fig.~\ref{fig:lrtest}.

Finally, we perform one more check, a variation on the Separated pair test, again using LRGs with line-of-sight separation $\Delta r_{\rm los} = 100 - 120 \mpch$.
The difference is that we now select the LRGs that overlap in projection, with separations $\rpair$ between 0.1 and $2\mpch$.
We then repeat the measurement as before, using the same estimator of filament signal.
Since the range of $\rpair$ does not match our other tests (or the filament measurement), it is not straightforward to directly plot the results of this test on Fig.~\ref{fig:lrtest} or \ref{fig:null-measure}.
Instead we simply check that the reduced $\chi^2$ statistic is consistent with the null hypothesis.
Indeed, with reduced $\chi^2 = 4 / 8$ the result is easily consistent with the null expectation.

\section{Discussion}
\label{sect:discussion}

We have presented a technique for the statistical measurement of properties of the dark matter filaments linking close pairs of galaxies and clusters.
Our method uses an empirical approach which relies on few assumptions to cancel out the contribution of spherical halos in the data.
Applying this technique to pairs of LRGs in SDSS we detect residual shear patterns of magnitude $\sim10^{-4}$, about one order of magnitude smaller than the LRG halo signal at 5 Mpc/h from the halo center, and two orders of magnitude smaller than the LRG signal at its virial radius. 
Employing a likelihood ratio test, we attribute this signal to filamentary structures with a detection significance higher than $4\sigma$.
The small signal, the possible systematics in SDSS imaging, and the dominance of the LRG halos make the measurement especially challenging.
We have carried out several null tests which provide evidence that our nulling technique is robust and that systematic errors are minimized.
Applying the same LRG pair selection and lensing estimator to simulations, we find an excellent match with the data.
With a width of a few $\mpch$ the stacked filament shape is determined to be thicker than the LRG halos at its end points.

There are several approximations and sources of error in our analysis. 
\begin{itemize}
\item The stacking of hundreds of thousands of LRG pairs leads to a smearing of the mass distribution. This means that we cannot make definitive statements about the typical (individual) filament structures in the universe, in particular the limits we obtain on the thickness of the filament only apply to the stacked profile. 
\item The binning scheme we use to null out the contribution of spherical halos also nulls {\it part} of the signal from the filaments we seek to measure. In this sense our estimator is sub-optimal (though it would be challenging to construct an optimal estimator while subtracting the LRG halo signal).
\item The calibration of the shear, which relies on a correction for the smearing due to the PSF, introduces a redshift dependent bias that propagates to the filament mass estimate. Uncertainties in the photometric redshifts of background galaxies have a similar effect. Both are below the 10 percent level and  smaller than our statistical uncertainties. 
\item Redshift space distortions: the line of sight separation of the LRGs is uncertain owing to their relative peculiar velocity. We have attempted to account for it by adding RSD to the simulations (see Sec.~\ref{sect:mocks}).
\item The inevitable contamination of the LRG sample with other galaxies and stars leads to a dilution of the signal. This should be controlled to better than the 10\% level. 
\end{itemize}

In future work several improvements can be made that address the above points. In addition, forward modeling of the measurement can be done using simulations and the halo model, so that comparisons can be made without use of our nulling technique. Such an approach may allow for more detailed tests of the halo model and of filamentary properties, though care will need to be exercised to distinguish systematic errors. Finally, an interesting complement to our study is to compare the mass distribution inferred from lensing shears with the distribution of foreground galaxies and hot gas. 

\section*{Acknowledgments}

We would like to thank Gary Bernstein, Sarah Bridle, J\"{o}rg Dietrich,
Mike Jarvis, Elisabeth Krause, Ravi Sheth, Yuanyuan Zhang and especially
Rachel Mandelbaum for helpful discussions and comments. We are very
grateful to Erin Sheldon for the use of his SDSS shear catalogs and to
Tomasz Kacprzak for related collaborative work.  BJ and MT also thank
the Aspen Center for Physics and NSF Grant \#1066293, for their warm
hospitality when part of this work was done. BJ and JC are partially
supported by Department of Energy grant DE-SC0007901.  MT is supported
by World Premier International Research Center Initiative (WPI
Initiative), MEXT, Japan, by the FIRST program ``Subaru Measurements of
Images and Redshifts (SuMIRe)'', CSTP, Japan, by Grant-in-Aid for
Scientific Research from the JSPS Promotion of Science (No.~23340061 and
26610058), and by MEXT Grant-in-Aid for Scientific Research on
Innovative Areas (No.~15H05893 and 15K21733). HM was supported in part by Japan Society
for the Promotion of Science (JSPS) Research Fellowships for Young
Scientists. HM was supported in part by the Jet Propulsion
Laboratory, California Institute of Technology, under a contract with
the National Aeronautics and Space Administration.

\appendix

\section{Halo ellipticity}
\label{app:ellip}

\begin{figure}
\centering
\resizebox{80mm}{!}{\includegraphics{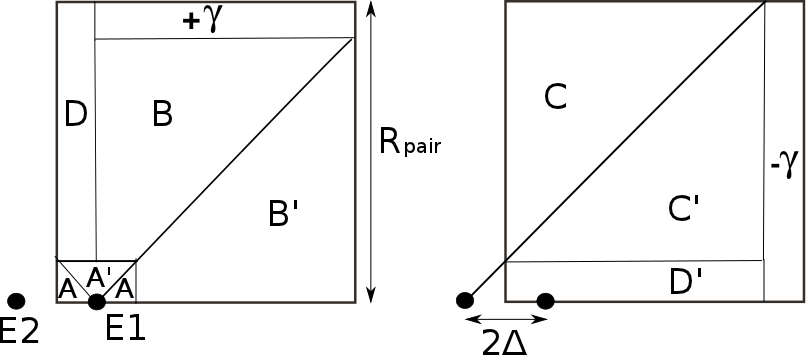}}
\caption{As an extreme model of halo ellipticity, we consider the shear from point masses E1 and E2. The two panels show the same region twice: the left panel highlights the contribution from E1, and the right that from E2. The net $\gamma_1$ shear (with the sign convention of Fig.~\ref{fig:points}) cancels in regions A and A', B and B', etc. (See the text for the details.) The size of the uncancelled regions, $+\gamma$ and $-\gamma$, shrinks rapidly with the small number $\Delta / \rpair \le 1/6$, showing that contributions from halo ellipticity are highly suppressed in our measurement.}
\label{fig:ellip}
\end{figure}

In order to show that the contribution from halo ellipticity is small, we consider a very simple model which is even less spherical than an elliptical halo. Thus, if the shear from this model is negligible, then so is shear from elliptical halos. We take two point masses labelled E1 and E2 on Fig.~\ref{fig:ellip}. These are each separated from the halo center by $\Delta \lesssim R_{\rm vir}$. The outermost square region pictured corresponds to the top square of Fig.~\ref{fig:points}, with side length $R_{\rm pair}$.

On the left panel of Fig.~\ref{fig:ellip} we extend two lines from E1 which are both 45 degrees from the horizontal axis. With our shear sign convention (Fig.~\ref{fig:points}), these lines describe points where the shear from E1 is purely $\gamma_2$, i.e., these lines are the zeros of $\gamma_1$. Thus, points which are on opposite sides of and equidistant from these lines have a net contribution of $\gamma_1 = 0$. As a result, the net $\gamma_1$ shear when summed over all galaxies in regions A and A' is zero. In the same way, regions B and B' sum to zero.

Likewise, on the right panel we draw a line from E2 which is 45 degrees from the vertical, and the net $\gamma_1$ shear in C and C' is zero. A final cancellation occurs in regions D and D', where the positive $\gamma_1$ shear from E1 in D cancels the negative shear from E2 in D'. The net shear from these two point masses is then given by the remaining regions, labelled $+\gamma$ and $-\gamma$. These two regions do not cancel perfectly, but it is clear that (i) these regions nearly cancel: while the $+\gamma$ region is slightly closer to E1 than the $-\gamma$ region is to E2, in area, the $+\gamma$ region is slightly smaller; (ii) the size of these imperfectly cancelled regions shrinks rapidly as $\Delta / \rpair$ gets smaller. The upper bound is
\be
\Delta / \rpair \le \frac{R_{\rm vir}}{\rpair} = \frac{1 \mpch}{6 \mpch} \, ,
\ee
but most of our LRG pairs have smaller virial radii and larger pair separation. Furthermore, the density profile of halos falls off quickly, so that relatively little of the mass is displaced an entire virial radius from the center.

\end{document}